\documentstyle[aps,preprint,tabularx,graphicx]{revtex}
\begin{document}
\draft
%
%
\title{Folding dynamics of the helical structure observed in a minimal model}
\author{Josh P. Kemp and Jeff Z.Y. Chen}
\address{Department of Physics,
University of Waterloo, Waterloo,
Ontario,Canada}
\date{\today}
\maketitle
\begin{abstract}
The folding of a polypeptide is
an example of the cooperative  effects of the amino-acid residues.
Of recent interest is how a secondary structure, such as a helix, spontaneously forms during
the collapse of a peptide from an initial denatured state. The Monte Carlo implementation
of a recent helix-forming model enables us to study the entire folding process dynamically.
 As shown by the computer simulations, the foldability and helical propagation are both strongly
correlated to the nucleation properties of the sequence.
\end{abstract}


\pagebreak
Attempting to understand the complex functional nature of proteins
is one of the most challenging problems in molecular biology. In
the past 10 years, considerable effort has been made to show that
these molecules are far from atypical polymer chains made of
disordered amino-acids. Despite the seemingly disordered nature of
the sequence, every protein possesses some remarkably similar
basic characteristics. Much of the current knowledge has been
derived from computer simulations of simplified protein
models\cite{sail,lau,go,sail1}. One of the ultimate goals of
theoretical modeling is to offer a base for quantitative
comparison with experimental structural determinations. Therefore,
it would be  advantageous to generalize minimal models in an
off-lattice, three-dimensional setting. For example, in an
off-lattice G\={o}-type model\cite{go}, a heteropolymer with
interactions between residues is constructed in such a way that
the interaction matrix is chosen to yield the desired native
state.\cite{go,shak,karplus}.

The essence of these lattice and off-lattice models is that
protein structures are created out of heterogeneity of the
sequence. There is no doubt that heterogeneity plays a dominant
role in structure selection; however, secondary structures are
known to originate from a number of other important effects such
as hydrogen bonding.  This suggests that additional considerations
should be made in theoretical models in order to capture
structural and dynamical properties that go beyond the
heterogeneity consideration in a sequence.

In a recent Letter, we have stressed the need to include a
directional biased residue-residue potential energy in order to
design a significantly ordered native state, using the helical
structure as an example\cite{kemp1}.  In particular, we have shown
that an almost perfect helical native structure could be produced
from a homopolymer backbone with a square-well potential that
prefers parallel bond angle planes (Fig.~\ref{dvec}). This
preference, written in a very compact mathematical form, can be
thought of as a simple adaptation of much more complex
hydrogen-bonding and dipole potential
energies\cite{kemp1,kollman,kemp2,kemp3}.

In this Letter, we present the folding dynamics study of
helix-forming polymers based on this model. In a typical numerical
experiment, a denatured initial configuration is well-equilibrated
at high temperature (Fig.~\ref{movie}a left), and is then quenched
below the coil-helix transition temperature at
$T=0.6\epsilon/k_B$, where $\epsilon$ is the maximum attractive
energy that two monomers can attain when forming a bond. A Monte
Carlo (MC) procedure is implemented after that point, and the
entire chain begins to collapse to the ordered helical state
(Fig.~\ref{movie}a right). The process allows us to examine the
kinetics of domain growth of the ordered segments.

The key features of the helical-forming model are kept the same as
in our previous study; however, some minor changes have been
implemented to tailor the model towards a more realistic helix
folding experiment. When randomly chosen, the rotation of a
monomer about the axis defined by the two nearest neighbors is
attempted according to the Metropolis rule\cite{metropolis}. The
bond angles can now fluctuate slightly around $2\pi/3$ with an
energy cost\cite{angle}, which maintains a worm-like  backbone and
allows for small local movements. The second change is the
replacement of the square-well potential for the monomer-monomer
interaction in our original work by the Lennard-Jones model which
allows for a more smooth dynamic motion when two monomers
interact\cite{LJ}. Our last change breaks the symmetry between
left and right handed helices. The vector $\hat u$ described in
Fig.~\ref{dvec} is tilted to make a correct right hand bonding in
a helical state after accounting for the pitch of the
helix\cite{uvec}.

In the first set of numerical experiments (Model I), chains with
number of residues $N=19, 25, 31, 37, 43,$ and $49$ were
considered. With each $N$, forty folding events were performed
with the allowed maximum folding time $t_{\rm max}$ listed in
Table 1.  Fig.~\ref{movie}a is a time lapse image of a typical
folding event ($N=49$), where the left plot corresponds to an
initial configuration, equilibrated at $T=\infty$,
and the last plot 
a completely folded state.
From the observation of these folding events, three important features emerge.

(1) For shorter chains ($N<31$) there is no clear indication of a
preferred nucleation site; the entire chain folds directly to the
ordered helix state. Almost all of the events that did not fold
into the preferred helical state acquire intermediate, poorly
wrapped globular states.

(2) For longer chains, the entire folding event accompanies a
nucleation propagation process, an expected mechanism for a
cooperative system.  As demonstrated in Fig.~\ref{movie}a, the
nucleation starts at the {\it ends} of the chain and gradually
heads towards the {\it center}, which is consistent with the
different mobility of residues along the chain. The terminal
residues are clearly more mobile due to reduced confinement
restrictions on the movement. Once the nucleation of the end of
the chain occurs, the helices begin to propagate inwards. At this
point, an interface between the two domains of helices might form.
This interface would eventually dissolve in favor of a single
uniform helix along the entire chain. In comparison with the
initial helical segmental formation which corresponds only to a
small fraction of the net folding time, resolving the
discontinuity is much slower and requires transverse fluctuations,
which are limited by the helical confining geometry.

(3) It is thus desirable to define order parameters which can be
used to quantitatively describe the kinetics of the helix
formation in our numerical experiments. For this model, the
$\hat{e}$ vector, defined relative to the vector $\hat{u}$ as
noted in Ref. 15 to correct for the pitch and the right handedness
of the helix, offers a unique direction for describing
orientational ordering. 
To study the local correlation of the bond orientation, we define the order parameter, $
H_1= (\sum_{i=1}^{N-1}{\hat e}_i\cdot{\hat e}_{i+1}) / ( {N-1})$.
%
%
%
In a chain with a final global helical structure, all the $\hat{e}$ vectors are more or less
 aligned so that this order parameter approaches unity. For a chain with helix domains
 as shown in some of the snapshots in Fig.~\ref{movie}a, this order parameter would also
yield a high value, although the two domains might have helical axes pointing in different
directions. Thus $H_1$ is a good measure of local helical content in a chain, but a poor measure of global ordering.
To study the global ordering in the system,  we define a second order parameter, $H_2
=\left|\sum_{i=1}^{N}{\hat e}_i\right| / N$.
%
%
This order parameter  deviates from unity
in fractured helix with opposite helical directions and approaches unity in a global helical state.

Typically, in folding studies of minimal models of proteins an
understanding of the folding properties can be obtained from a
study of the so-called ``first passage''
times\cite{sail,sail1,gutin}, defined as the time required for a
molecule to first enter the native state when started in an
arbitrary configuration. Here, when  both order parameters reach a
value greater than $0.95$ the segment is regarded as helical,
specified to ensure that the segments are near perfect helices.
The mean first passage times (MFPTs) 
shown in Table 1.

Good folding proteins have MFPTs that obey a power-law behavior when scaled with system size\cite{gutin}.
%
\begin{equation}
\label{power}
t_{\rm mfp}\sim N^{\lambda}~,
\end{equation}
%
where $t_{\rm mfp}$ is the averaged first passage time, and
$\lambda$ is a characteristic exponent. The exponent $\lambda$
varies depending on how the sequences were designed. For example,
random sequenced
 chains scaled with $\lambda\approx 6$,
and sequences designed from a Miyazawa and Jernigan\cite{mj}
potential scaled with $\lambda= 4.5$\cite{gutin}.  It has been
observed that sequences designed from protein-like potentials were
better folders (had smaller $\lambda$).

Using the data from Table 1, we can determine $\lambda$ associated
with our helical model by examining the data on a log-log plot
(Fig.~\ref{fig1}). Fitting the data with a least-square method to
Eq.~\ref{power} we obtained a value of $\lambda_{\rm 1}=3.7(2)$.
Our model demonstrates the characteristics of a well designed
protein sequence, not a surprising conclusion since we know that
helices exist in real proteins. What makes a crucial difference in
comparison with previous results is that we are dealing with a
{\it homopolymer} here, not a hetropolymer in other studies.

The fact that helix nucleation starts from the terminals rather than the center leads to a
simple question: can the folding scenario of the entire chain be altered by designing a heteropolymer
that contains monomers with different functionality? In particular, as our
second set of numerical experiments,  two segments of six monomers were attached to the original
 terminals the helix-forming chain (Model II). These new segments are neutral and only interact through an excluded volume interaction {\it with no attractions}. The native structure of the new chain is completely determined by the helical-forming segment,
and the neutral terminal segments only display a partial random coil conformation.
 Now, the terminal residues of the attractive segment
 no longer have the high mobility and will have nearly the same likelihood of nucleation as the interior monomers.  With the addition of the non-attractive segments, the chain is now considered helical if the attractive residues, not the added ones, meet the requirements stated above for helicity.

A typical folding event is displayed in Fig.~\ref{movie}b in a series of time lapse plots,
with the neutral residues represented in black. In contrast to our previous set of experiments,
it is now more likely to form a single, central helical nucleation site,
rather than the two-domain structure that we observed before.
It would appear that adding the two neutral sections has a net effect of slowing down the dynamics of the segment,
as we know that longer segment will have much slower dynamic response. However, a striking feature of adding neutral
segments is that the MFPTs actually {\it decrease} dramatically in comparison with its isolated counterpart,
as displayed in  Table~\ref{data2}. 
The reduction in ability of the terminal residues to nucleate causes a more
uniform distribution of nucleation sites along the chain, and
decreases the overall nucleation probability.
This means that the initial nucleation is longer, but a nucleation site that
already exists has a much longer time to propagate through the entire segment before a second nucleation site occurs.
Thus, there is a significant reduction in the folding times because a discontinuity
in the segment does not have to be resolved. 
Multiple nucleation sites can still occur, 
 however, 
only $50\%$ of chains can fold with a single nucleation region. In contrast, an isolated helical segment almost always folds with a discontinuity.

We have made a  log-log plot of the data in Fig.~\ref{fig1},
where the MFPTs are fitted to the same power law  in Eq.~\ref{power}. The characteristic exponent
$\lambda_{\rm 2}=2.4(3)$ is
significantly lower than the exponent found above ($\lambda_1=3.7$).
 The new $\lambda_2$ demonstrates that the folding process is fundamentally different. 
Longer helices are more easily formed if the probability of seeding a segment is relatively small
 compared to the propagation time of the isolated helical segment.

To observe the folding kinetics from yet another angle, we examine a third model (Model III)
in which we attach only a single neutral segment to one end of the helix-forming chain.
The folding times for this model are shown in Table~\ref{data2} 
and the fitted
$\lambda_{\rm 3}=3.5(1.0)$.
The dynamics of this type of segment is a combination of the two segments already discussed.
 The nucleation of the helical segment occurs at the free end of the chain,  as in an isolated segment.
Now, what differs is that this is likely to be the only nucleation site,
thus reducing the probability of having to resolve a discontinuity;
the propagation of the helical segment occurs through longitudinal fluctuations
 along the chain contour and is retarded by sharp changes in the chain contour.
This slowing of the propagation  provides an opportunity to generate a second nucleation
site in the remaining segment which might give
 rise to a discontinuity that retards the dynamics.
Only approximately $15\%$ of the chains now fold with a single nucleation site.

Thus,  $\lambda$  is
sensitive to the probability of multiple nucleation sites.
More than one nucleation site decrease the foldability of the segment by creating a discontinuity.
In our previous work\cite{kemp3}, the anisotropy of the potential was
demonstrated
 to play a significant role in the foldability of a helical segment.
Klimov and Thirumalai\cite{klimov} postulated that the foldability of a protein is related to
 the relative separations of the coil-globular and globular-folded transition through the parameter,
 $\sigma={{T_f-T_{\theta}}/{T_{\theta}}}$,
%
%
where $T_{\theta}$ is the coil-globular transition
 temperature
and $T_f$ the globular-helix transition
temperature.
 We showed that decreasing the anisotropy increases the value of $\sigma$ in our model.

To further explore the kinetc features of Model II above where $m=6$ was used (high anisotropy),
we design a new set of experiments using
$m=2$ (low anisotropy)
The MFPTs were collected for the same helical segment lengths as shown in Table 1,
with a $\lambda_2(m=2)=3.1(5)$ which is higher than $\lambda_2(m=6)=2.4$.
Comparing the data in Table 1 as well
 as the corresponding $\lambda$'s we conclude that there is indeed
a reduction in the folding times due to the reduction in anisotropy. 
 Results from our pervious work show that there is a globular state at high temperatures
 which becomes stable as the anisotropy is decreased. This increased stability will reduce
the probability of nucleation and decrease the rate of helical propagation, thus accounts
for the observed increase in 
$\lambda$. 


In summary, we
have shown for the first time that the folding times from the coil state to the helical state scales
as a power-law with the system size, as expected for protein-like system. 
Both folding times and the scaling exponents for the system are altered as well when
  the nucleation probability is adjusted. 
Nucleation is not the only important factor in folding, and the rate of propagation to nucleation is a
dominating factor in fast folding characteristics of a helical segment.
    These results demonstrate the significance of ``hot'' sites, or conserved residues\cite{hot},
 within a protein. These sites are the key nucleation regions of the folding process and important
for the creation the native state. 
 As we have demonstrated
in these simulations, it is important to creat a dominate nucleation to ensure that
 propagation can proceed  throughout the entire chain without an alternate nucleation site
forming; multiple ``hot'' sites would prolong the folding time if they are formed too early in the folding.

We would like to thank NSERC for the financial support of this work.



\begin{figure}
\caption{Interaction between two residues labeled $i$ and $j$. The ${\vec u}$ vector is
nomal to the bond plan. A modified Lennard-Jones
interaction is assumed with the modification that the vector $\hat u_i$ prefers to align with
the vector $\hat u_j$ (see definition in footnote 19).}
\label{dvec}
\end{figure}

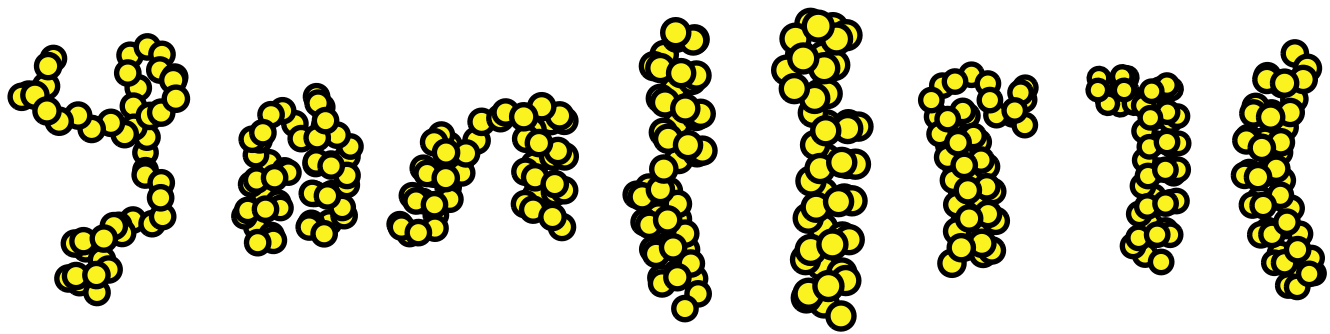
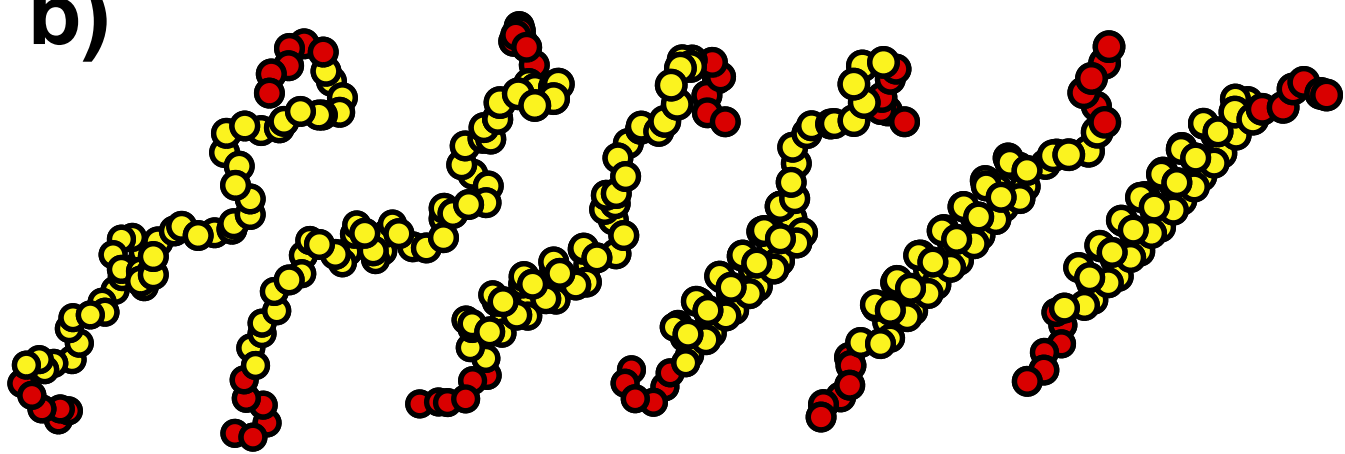
\begin{figure}
\caption{Typical folding scenarios for a helical segment of length $N=49$:
(a) Multiple nucleation and (b)  single nucleation }.
\label{movie}
\end{figure}

\begin{figure}
\caption{Scaling of average folding time vs. polymer length, for
helical
segments (squres) and helical segments with two tethered segments (circles).}
\label{fig1}
\end{figure}


\begin{table}
\caption{Folding time table. $N$ is the number of helix forming residues,
$t_{\rm mfp}$ is the average first passage time, %
and
 \% DNF, is the percentage that did not fold. In model I, the maximum folding time allowed
is 10, 20, 50, 100, 150, 200 ($\times 10^6$MC steps) and in model
II and III, 25, 50, 100, 150, 200, 250  ($\times 10^6$MC steps) ,
for $N=19,25,31,37,43,49$, respectively.}
\begin{center}
\begin{tabular}{|@{\hspace{0.25cm}}c@{\hspace{0.25cm}} |@{\hspace{0.25cm}}c@{\hspace{0.5cm}}c@{\hspace{0.25cm}}
|@{\hspace{0.25cm}}c@{\hspace{0.5cm}}c@{\hspace{0.25cm}}
|@{\hspace{0.25cm}}c@{\hspace{0.5cm}}c@{\hspace{0.25cm}}
|@{\hspace{0.25cm}}c@{\hspace{0.5cm}}c@{\hspace{0.5cm }}|}
\hline
 { } & \multicolumn{2}{c} {Model I (m=6)} & \multicolumn{2}{c}{Model II (m=6)}&\multicolumn{2}{c}{Model III (m=6)}
&\multicolumn{2}{c}{Model II (m=2)}\\\hline $N$  &$t_{\rm
mfp}(\times 10^6)$&\%DNF &$t_{\rm mfp}(\times 10^6)$ &\%DNF
&$t_{\rm mfp}(\times 10^6)$  &  \%
DNF&$t_{\rm mfp}(\times 10^6)$ &  \% DNF\\
\hline
19 & 1.2(2) & 33& 4.5(5) & 33& 3.0(5) & 20 &2.4(4) &40\\
25 & 4.3(6) & 10& 6.7(9) & 20 & 7.0(8) & 50 &6.6(2) & 35\\
31 & 12(2)  & 10 & 13.(3) & 25& 10.(2) &35 &14.(6) &30\\
37 & 26(4)  & 10 & 16.(3) & 23& 39.(5) & 40 &29.(10) &40\\
43 & 50(7)  & 20 & 33.(6) & 15 & 28.(5) &15 &59.(16) &40\\
49 & 72(8)  & 15 & 36.(8) & 13 & 73.(14) & 10 &69.(18) &35\\
\hline
\end{tabular}
\end{center}
\label{data2}
\end{table}






\begin{references}
\bibitem{ran} E.I. Shakhnovich, A.M. Gutin, {\it Nature} {\bf 346}, 773 (1990)
\bibitem{sail}  A. \u{S}ail, E. Shakhnovich, M. Karplus, {\it Nature} {\bf 369}, 248, (1994)
\bibitem{lau} K.F. Lau, K.A. Dill, {\it Macromolecules} {\bf 22}, 3986 (1989)
\bibitem{go}H. Taketomi, Y. Ueda, N. G\={o}, {\it Int. J. Peptide Protein Res.} {\bf 7}, 445 (1975); N. G\={o} and H. Abe, Biopolymers {\bf 20}, 991 (1981); N. G\={o} and H. Abe, Biopolymers {\bf 20}, 1013 (1981)
\bibitem{sail1} A. \u{S}ail, E. Shakhnovich, M. Karplus, {\it J. Mol. Biol.} {\bf 235}, 1614 (1994)
\bibitem{shak} N.V. Dokholyan, S.V. Buldyrev, H. E. Stanley, E.I. Shakhnovich, {\it cond-mat/9812291} (1998); N.V. Dokholyan, S.V. Buldyrev, H. E. Stanley, E.I. Shakhnovich, {\it cond-mat/9812284} (1998)
\bibitem{karplus} Y. Zhou, M. Karplus, {\it Proc. Natl. Acad. Sci. USA} {\bf 94}, 14429 (1997)
\bibitem{kemp1}J.P Kemp and Z.Y. Chen, {\it Phys. Rev. Lett.} {\bf 81}, 3880(1998)
\bibitem{kollman}P. A. Kollman and S.J. Weiner, {\it J. Comp. Chem.} {\bf 2}, 287 (1981); S.J Weiner {\it et al.}, {\it J. Am. Chem. Soc.} {\bf 106}, 765 (1984)
\bibitem{kemp2}J.P. Kemp, U.H.E. Hansmann, Z.Y. Chen, {\it Eur. Phys. J. B} {\bf 15} 371(2000)
\bibitem{kemp3}J.P. Kemp, J.Z.Y. Chen, {\it Biomacromolecules}, in press.
\bibitem{metropolis} N. Metropolis, A.W. Rosenbluth, M. N. Rosenbluth, A. H. Teller, E. Teller, {\it J. Chem. Phys.} {\bf 21} 1087 (1953)
\bibitem{angle} The exact expression for the potential energy of  a bond angle $\theta$ is given by
$K ({\rm cos} \theta + {\sqrt 3}/2)^2$
where $K = 10 \epsilon $ in our simulation.

\bibitem{LJ}The Lennard-Jones potential is used for the attractive part of
monomer-monomer interaction when two monomers have a distance $r$,
%
\begin{equation}
\label{ljpotential}
V(r) = -\epsilon\left({{a_0}\over{r^{12}}} -
{{a_1}\over{r^6}}\right)~,
\end{equation}
%
where the parameters $a_0$ and $a_1$ are chosen such that $V(d)=0$, where $d=(3/2)a$ with $a$ being the bond length.

\bibitem{uvec} This way the $\hat{e}$ vector points directly along the helical axis when
a perfect helical structure is formed. The vectors are defined as follows,
%
$ {\hat u}_i =  ({\vec r}_{i+1}-{\vec r}_i)\times ({\vec r}_i -
{\vec r}_{i-1}) / |({\vec r}_{i+1}-{\vec r}_i)\times ({\vec r}_i -
{\vec r}_{i-1}) | $ and $ {\hat e}_i = \sqrt{1-D^2}{\hat u}_i + D
({\vec r}_{i+1} - {\vec r}_{i-1}) / |({\vec r}_{i+1} - {\vec
r}_{i-1})| $ where $D$ is the separation of sub-units along the
helical axis, such that the pitch is $p=nd$ given that there are
$n$ monomers per loop. The complete form of the potential energy
is
\begin{equation}
\label{potential}
U({\vec r})=\left\lbrace\begin{array}{cl} 
V(r)\left[ ({\hat e}_i\cdot {\hat r}_{ij})\cdot ({\hat e}_i\cdot {\hat r}_{ij}) \right]^m & for\ d\leq r_{ij} < \sigma\\
\infty & for\ 0\leq r_{ij} < d \\
\end{array}
\right.
\end{equation}
where ${\hat r}_{ij}=({\vec r}_i-{\vec r}_j)/|{\vec r}_i-{\vec
r}_j|$. The parameter $m$ controls the anisotropy of the potential
and is set to $m=6$.
The  adjustable parameter $\epsilon$ is scaled  into the
temperature to produce the reduced temperature unit
$\tilde{T}=k_BT/\epsilon$.
\bibitem{gutin} A.M. Gutin, V.I. Abkevich, E.I. Shakhnovich, {\it Phys. Rev. Lett.} {\bf 77}, 5433 (1996)
\bibitem{mj}S. Miyazawa and R.L. Jernigan, {\it Macromolecules} {\bf 18}, 534 (1985)
\bibitem{hot} E.I. Shakhnovich, V. Abkevich, O. Ptitsyn, {\it Nature} {\bf 379}, 96 (1996); M. Skoroboagatiy and G. Tiana, {\it Phys. Rev. E} {\bf 58}, 3572 (1998)
\bibitem{klimov} D.K. Kilmov, D. Thirumalai, {\it Folding and Design} {\bf 3}, 127 (1998)

%

\end{references}
\end{document}